\documentclass{article}

\usepackage{float}
\usepackage[numbers]{natbib}
\usepackage{amssymb}
\usepackage{amsmath}
\usepackage{booktabs}
\usepackage{graphicx}

\title{A Novel Multi-view Mixture Model Framework for Longitudinal Clustering with Application to ANCA-Associated Vasculitis}

\author{
Shen Jia$^{1,2}$\thanks{Corresponding author: shjia@tcd.ie}, 
David Selby$^{3}$, 
Mark A Little$^{2,4}$, 
Tin Lok James Ng$^{1,2}$ \\
\\
\small $^{1}$School of Computer Science and Statistics, Trinity College Dublin, Dublin, Ireland \\
\small $^{2}$ADAPT Center, Dublin, Ireland \\
\small $^{3}$German Research Centre for Artificial Intelligence (DFKI), Kaiserslautern, Germany \\
\small $^{4}$Trinity Kidney Centre, Trinity College Dublin, Dublin, Ireland
}
\begin{document}

\maketitle

\begin{abstract}
Effectively modeling irregularly sampled longitudinal data is essential for understanding disease progression and improving risk prediction. We propose a two-view mixture model that integrates static baseline covariates and  longitudinal biomarker trajectories within a unified probabilistic clustering framework. Temporal patterns are modeled using Neural Ordinary Differential Equations. Model training uses an EM algorithm with a sparsity-inducing log-penalty for interpretable subgroup discovery. Application of the model to an Irish cohort of ANCA-associated vasculitis patients reveals subgroups with heterogeneous serum creatinine trajectories and variation in end-stage kidney disease outcomes. 
\end{abstract}


\section{Introduction}

ANCA-associated vasculitides (AAV) are rare autoimmune disorders affecting small- to medium-sized blood vessels, with predominant involvement of the kidneys and lungs \cite{Jennette2014, Kallenberg2014,carney2024urineimmune}. Although outcomes have improved with modern immunosuppressive therapies \cite{Mukhtyar2009, Harper2005}, progression to end-stage kidney disease (ESKD) remains a major clinical concern, particularly among patients with severe renal involvement at diagnosis. Conventional monitoring tools—including ANCA titers and standard renal biomarkers such as serum creatinine and estimated glomerular filtration rate (eGFR)—are useful for assessing kidney function but have limited ability to capture subtle early deterioration \cite{Inker2014}. Recent evidence suggests that analyzing longitudinal trajectories of serum creatinine may provide earlier insight into renal decline and ESKD risk, even when absolute values remain within normal ranges \cite{Hanberg2023renalpatterns, Kim2023}. This highlights the importance of dynamic, time-dependent modeling approaches for improving risk stratification and early identification of patients at heightened risk for progressive kidney failure.

Patients with AAV exhibit substantial clinical heterogeneity in disease presentation and progression. Identifying latent subgroups through clustering is critical for uncovering distinct patterns of disease evolution, guiding personalized treatment, and improving prognosis. While various methods jointly model static and longitudinal data in supervised contexts, integration in an unsupervised clustering framework remains rare. Existing clustering approaches often either ignore temporal dynamics or reduce longitudinal data to summary statistics, limiting their ability to capture complex trajectory patterns.

Multi-view clustering  has become a powerful approach for integrating heterogeneous data sources that provide complementary information about the same entities \cite{gao2019are, chao2017survey}. Existing methods range from unified probabilistic models to deep learning–based fusion techniques and have been successfully applied in fields such as neuroimaging \cite{Feng2021} and electronic health records (EHRs) \cite{Ho2020}. These approaches integrate diverse data modalities, including demographics, laboratory tests, and imaging, to uncover latent subgroups.

However, most multi-view clustering methods do not readily accommodate irregularly sampled longitudinal processes \cite{gao2019are, chao2017survey, zhou2018}, which are common in clinical settings. Integrating static baseline features with longitudinal biomarker trajectories poses two key challenges: (i) the fundamentally different statistical nature of static versus temporal data, and (ii) the sparsity and irregular timing of clinical time series, which are poorly addressed by traditional feature engineering or simple discretizations.

Mathematically, the problem can be formulated as follows: we observe data points \(\{x_i\}_{i=1}^N\), where each observation \(x_i\) consists of two views, \(x_i = (x_i^{(1)}, x_i^{(2)})\). The first view, \(x_i^{(1)} \in \mathbb{R}^{d^{(1)}}\), is a fixed-dimensional feature vector. The second view, \(x_i^{(2)} = \{x_{i,j}^{(2)}\}_{j=1}^{n_i}\), is a set of longitudinal measurements taken at individual-specific time points \(\{t_{i,j}\}_{j=1}^{n_i}\). Importantly, both the number of measurements \(n_i\) and the timing \(\{t_{i,j}\}\) vary across subjects. The goal of multiview mixture modeling is to define a probabilistic framework that jointly leverages both views to uncover latent subgroups, capturing complementary information from static and longitudinal data in a unified manner.

We propose a novel multiview mixture model that jointly clusters patients by integrating fixed-dimensional baseline features with continuous-time trajectories. A key innovation of our approach is the use of neural ordinary differential equations (Neural ODEs) to model longitudinal trajectories \cite{ chen2018neural}. Recent developments in continuous-time latent variable modeling, such as Neural ODEs, allow for the direct learning of smooth latent trajectories from irregularly sampled clinical data. By unifying static and longitudinal data within a probabilistic framework, our method uncovers clinically meaningful subgroups associated with relapse risk and renal outcomes, facilitating more personalized management strategies in AAV. Our proposed model is broadly applicable well beyond the AAV context. Indeed, many biomedical and other applications involve data consisting of both fixed-dimensional features and longitudinal measurements.

The remainder of the manuscript is organized as follows. Section \ref{sec_background} provides background on Neural ODEs and multiview mixture models. In Section \ref{sec_method}, we present our proposed multi-view mixture model for the joint analysis of static and longitudinal data, along with an Expectation-Maximization algorithm for parameter estimation. Section \ref{sec_simulation} reports the results of simulation studies, and Section \ref{sec_data_app} illustrates the application of the proposed model to patients in Ireland diagnosed with antineutrophil cytoplasmic antibody –associated vasculitis. Finally, Section \ref{sec_discussion} discusses potential directions for future work.

\section{Background}
\label{sec_background}
\subsection{Neural Ordinary Differential Equation}
\label{sec_neural_ode}
A Neural ODE (Neural Ordinary Differential Equation) \citep{chen2018neural} is a way to model continuous-time dynamics using neural networks. 
Let \( z: [0, T] \to \mathbb{R} \) represent the hidden state evolving over time. A Neural ODE models the evolution of this hidden state by defining its time derivative as:
\begin{equation}\label{eqn_neural_ode_diff_eq}
\frac{dz(t)}{dt} = f_{\theta}(z(t), t) .
\end{equation}
Here, \( f_{\theta}: \mathbb{R}^2 \rightarrow \mathbb{R} \) is a neural network parameterized by \( \theta \) that represents the dynamics of $z$. It takes as input the current hidden state \( z(t) \) and time \( t \), and outputs the corresponding derivative at $t$. Thus, to compute the hidden state $z(t_1)$ at time $t_1$ given hidden state $z(t_0)$ at time $t_0$, we solve an ODE
\begin{equation}
    z(t_1) = z(t_0) + \int_{t_0}^{t_1} f_\theta(z(t), t) dt .
\end{equation}
Given the function \( f_{\theta} \) and an initial condition \( z(t_0) = z_{t_0} \) at time \( t_0 \), the hidden state at any set of time points can be computed using numerical ODE solvers, such as Runge--Kutta methods:
\begin{equation}\label{eqn_nerual_ode_solver}
 z(t_1), \dots, z(t_m) = \text{ODESolve}(f_{\theta}, z_{t_0}, (t_0, t_1, \dots, t_m))
\end{equation}
where \( t_1, \dots, t_m \) are arbitrary time points.

There are several possible architectures for \( f_{\theta} \), with the feedforward neural network being the simplest and most widely used. In this work, we focus on using a feedforward network for \( f_{\theta} \) with two hidden layers with width $h_1$ and $h_2$ respectively. Mathematically, \( f_\theta \) can be expressed as
\begin{equation}
\label{eqn_feedforward_nn}
f_\theta(z(t), t) = W^{(3)} \, \sigma \left( W^{(2)} \, \sigma \left( W^{(1)} [z(t), t]^{T} + \mathbf{b}^{(1)} \right) + \mathbf{b}^{(2)} \right) + b^{(3)},    
\end{equation}

where \( \theta \) consists of the weights and biases of the network: \( \theta = \{ W^{(1)}, \mathbf{b}^{(1)}, W^{(2)}, \mathbf{b}^{(2)}, W^{(3)}, b^{(3)} \} \). Here, \( W^{(1)} \in \mathbb{R}^{h_1 \times 2} \), \( \mathbf{b}^{(1)} \in \mathbb{R}^{h_1} \), \( W^{(2)} \in \mathbb{R}^{h_2 \times h_1} \), \( \mathbf{b}^{(2)} \in \mathbb{R}^{h_2} \), \( W^{(3)} \in \mathbb{R}^{1 \times h_2} \), and \( b^{(3)} \in \mathbb{R} \). The activation function \( \sigma(\cdot) \) is the element-wise sigmoid, defined for a scalar \( x \) as
\[
\sigma(x) = \frac{1}{1 + e^{-x}},
\]
and applied component-wise to a vector \( \mathbf{x} \in \mathbb{R}^d \) as
\[
\sigma(\mathbf{x}) = (\sigma(x_1), \dots, \sigma(x_d)),
\]
mapping each element to the interval \( (0,1) \).
We choose a feedforward network with two hidden layers for $f_\theta$ as it provides sufficient capacity to model complex nonlinear dynamics while keeping the network architecture simple and computationally efficient.

Neural ODEs offer distinct advantages for modeling the dynamics of hidden states. In particular, they naturally support continuous-time modeling, making them well suited for irregularly sampled time series, such as longitudinal data. The dynamics are governed by the neural network \( f_\theta \), and by adjusting its complexity, for example by increasing the number of hidden units, it can approximate a broad class of functions, thanks to the universal approximation theorem for neural networks \citep{Hornik1989}.

\subsubsection{Neural ODE for Longitudinal Data}
Neural ODEs are a natural choice for modeling the dynamics of longitudinal data. Suppose we are given a longitudinal dataset consisting of \( N \) individuals, with data points \( x_1^{(2)}, \ldots, x_N^{(2)} \). For each individual \( i \), the observations are denoted by \( x_i^{(2)} = \{x^{(2)}_{i,1}, \ldots, x^{(2)}_{i,n_i}\} \), recorded at subject-specific time points \( t_i = \{t_{i,1}, \ldots, t_{i,n_i}\} \). That is, \( x_{i,j}^{(2)} \) represents the measurement for individual \( i \) taken at time \( t_{i,j} \). Here, \( n_i \) represents the number of time points (or encounters) for individual \( i \), which may differ across subjects.

Following the approach introduced by \cite{chen2018neural}, the key idea is to assume that the underlying mean trajectory or overall pattern of evolution is shared across the \( N \) individuals in the longitudinal dataset. This shared dynamic is modeled using a neural ODE. More concretely, let \( z: [0,T] \rightarrow \mathbb{R} \) denote the latent trajectory, which is modeled by the neural ODE \eqref{eqn_neural_ode_diff_eq} with neural network \( f_\theta \). For each individual \( i \), the observation \( x^{(2)}_{i,j} \) at time \( t_{i,j} \) is assumed to be normally distributed with mean given by the latent trajectory \( z(t_{i,j}) \) at that time and variance \( \sigma^2 \):
\begin{equation}
\label{eqn_normal_dist_longitudinal}
x^{(2)}_{i,j} \sim \mathcal{N}(z(t_{i,j}), \sigma^2).    
\end{equation}
In other words, the latent trajectory \( z(t) \) represents the longitudinal mean across the sample of \( N \) individuals.

\subsection{Multi-view Mixture Model}
\label{sec_multiview_mixture}
We consider a multi-view mixture model with \( V \geq 2 \) views. In this framework, each view is assumed to follow a marginal mixture distribution, and, conditional on the cluster memberships, the views are independent. For view \( v = 1, \ldots, V \), let \( x^{(v)} \in \mathbb{R}^{d^{(v)}} \) represent the feature vector associated with the \( v \)-th view, which contains \( K^{(v)} \) distinct clusters. The latent cluster membership for view \( v \) is denoted by \( \xi^{(v)} \in \{1, \ldots, K^{(v)}\} \). 

Let \( \phi^{(v)}(x^{(v)} \mid \varphi_k^{(v)}) \) denote the conditional density of the \( k \)-th cluster in the \( v \)-th view, parameterized by \( \varphi_k^{(v)} \). The joint cluster membership probabilities across all views are captured by the tensor \( \pi \in \mathbb{R}^{K^{(1)} \times \cdots \times K^{(V)}} \), whose non-negative entries sum to 1. Specifically, each entry 
\begin{equation}
\label{eqn_cluster_prob_tensor}
\pi_{k^{(1)}, \ldots, k^{(V)}} := P(\xi^{(1)} = k^{(1)}, \cdots, \xi^{(V)} = k^{(V)}) .     
\end{equation}
represents the probability that an observation simultaneously belongs to cluster $k^{(v)}$ in each view. Marginal mixture probabilities for a given view $v$ can be obtained by summing over the other views:

\begin{equation}
\label{eqn_cluster_prob_marginal}
\pi^{(v)}_k := P(\xi^{(v)} = k) = 
\sum_{k^{(1)}=1}^{K^{(1)}}
\cdots
\sum_{k^{(v-1)}=1}^{K^{(v-1)}}
\sum_{k^{(v+1)}=1}^{K^{(v+1)}}
\cdots
\sum_{k^{(V)}=1}^{K^{(V)}}
\pi_{k^{(1)}, \dots, k^{(v-1)}, k, k^{(v+1)}, \dots, k^{(V)}}
\end{equation}
    
Consider a sample of $N$ observations, each with $V$ views, represented as $\{ x_i \}_{i=1}^N$, where $x_i = (x_i^{(1)}, \ldots, x_i^{(V)})$ and $x_i^{(v)} \in \mathbb{R}^{d^{(v)}}$. Given the numbers of view-specific clusters $K^{(1)}, \ldots, K^{(V)}$, the log-likelihood function is:

\begin{equation}
\label{eqn_log_likelihod_multiview}
    \ell(\{x_i\}_{i=1}^{N} \mid \boldsymbol{\varphi}, \pi) = \sum_{i=1}^{N} \log \left( \sum_{k^{(1)}=1}^{K^{(1)}} \cdots \sum_{k^{(V)}=1}^{K^{(V)}} 
    \pi_{k^{(1)}, \dots, k^{(V)}} 
    \prod_{v=1}^V \phi^{(v)}\left(x_i^{(v)} \mid \varphi^{(v)}_{k^{(v)}}\right) \right),
\end{equation}

where \( \boldsymbol{\varphi} := \{ \varphi_{k^{(v)}}^{(v)} \} \) denotes the set of view- and cluster-specific parameters for all \( v = 1, \ldots, V \) and \( k^{(v)} = 1, \ldots, K^{(v)} \).

\subsubsection{EM Algorithm}
\label{sec_background_em}
Similar to the single-view mixture model, the Expectation--Maximization (EM) algorithm \citep{Dempster1977} can be used to maximize the log-likelihood in Equation~\eqref{eqn_log_likelihod_multiview}. Given the current parameter estimates $(\boldsymbol{\varphi}^s, \pi^s)$, the E-step constructs the expected complete-data log-likelihood:
\begin{equation}
\label{eqn_multiview_complete_data_ll}
Q(\boldsymbol{\varphi}, \pi \mid \boldsymbol{\varphi}^s, \pi^s)
=
\sum_{i=1}^{N}
\sum_{k^{(1)}=1}^{K^{(1)}} \cdots \sum_{k^{(V)}=1}^{K^{(V)}}
\gamma^s(k^{(1)}, \ldots, k^{(V)} \mid x_i)
\log \left(
\pi_{k^{(1)}, \dots, k^{(V)}}
\prod_{v=1}^V
\phi^{(v)}\!\left(x_i^{(v)} \mid \varphi^{(v)}_{k^{(v)}}\right)
\right),
\end{equation}
where
\begin{equation}
\label{eqn_multiview_cluster_assign_prob}
\gamma^s(k^{(1)}, \ldots, k^{(V)} \mid x_i)
=
\frac{
\pi^s_{k^{(1)}, \dots, k^{(V)}}
\prod_{v=1}^V
\phi^{(v)}\!\left(x_i^{(v)} \mid \varphi^{(v),s}_{k^{(v)}}\right)
}{
\sum_{j^{(1)}=1}^{K^{(1)}} \cdots \sum_{j^{(V)}=1}^{K^{(V)}}
\pi^s_{j^{(1)}, \ldots, j^{(V)}}
\prod_{v=1}^V
\phi^{(v)}\!\left(x_i^{(v)} \mid \varphi^{(v),s}_{j^{(v)}}\right)
}
\end{equation}
denotes the posterior probability of the joint cluster assignment for observation $x_i$ under the current parameter estimates.

In the M-step, the parameters are updated by maximizing the expected complete-data log-likelihood:
\begin{equation}
(\boldsymbol{\varphi}^{s+1}, \pi^{s+1})
=
\arg\max_{\boldsymbol{\varphi}, \pi}
Q(\boldsymbol{\varphi}, \pi \mid \boldsymbol{\varphi}^s, \pi^s).
\end{equation}
From Equation~\eqref{eqn_multiview_complete_data_ll}, the optimization problem is separable with respect to $\pi$ and $\boldsymbol{\varphi}$. In particular, the update for the joint cluster probability tensor $\pi$ admits a closed-form solution:
\begin{equation}
\pi^{s+1}_{k^{(1)}, \dots, k^{(V)}}
=
\frac{1}{N}
\sum_{i=1}^{N}
\gamma^s(k^{(1)}, \dots, k^{(V)} \mid x_i).
\end{equation}
Using the marginalization relationship in Equation~\eqref{eqn_cluster_prob_marginal}, the view-specific marginal mixture probabilities
$\{ \pi_k^{(v), s+1} \}_{k=1}^{K^{(v)}}$ can be obtained for each view $v$.

Finally, for each view $v$ and each corresponding cluster $k = 1, \ldots, K^{(v)}$, the cluster-specific parameters $\varphi_k^{(v)}$ are updated by solving:
\begin{equation}
\label{eqn_multiview_m_step_clus_para}
\varphi_k^{(v), s+1}
=
\arg\max_{\varphi_k^{(v)}}
\sum_{i=1}^{N}
\sum_{k^{(1)}=1}^{K^{(1)}} \cdots
\sum_{k^{(v-1)}=1}^{K^{(v-1)}}
\sum_{k^{(v+1)}=1}^{K^{(v+1)}} \cdots
\sum_{k^{(V)}=1}^{K^{(V)}}
\gamma^s\!\left(
k^{(1)}, \ldots, k^{(v-1)}, k, k^{(v+1)}, \ldots, k^{(V)}
\mid x_i
\right)
\log \phi^{(v)}\!\left(x_i^{(v)} \mid \varphi_k^{(v)}\right).
\end{equation}

\subsubsection{Sparsity Inducing Log Penalty}
\label{sec_background_sparsity}
The cluster membership probability tensor \( \pi \) in \eqref{eqn_cluster_prob_tensor} encodes how information is shared across the \( V \) views. Specifically, an entry \( \pi_{k^{(1)}, k^{(2)}, \ldots, k^{(V)}} = 0 \) indicates that an observation cannot simultaneously belong to cluster \( k^{(1)} \) in view 1, cluster \( k^{(2)} \) in view 2, and so on, up to cluster \( k^{(V)} \) in view \( V \). In many real-world applications, the tensor \( \pi \) is expected to be sparse, with many entries equal to or near zero. 

Recall that the M-step of the EM algorithm involves maximizing the expected complete-data log-likelihood \( Q(\boldsymbol{\varphi}, \pi \mid \boldsymbol{\varphi}^s, \pi^s) \) as given in \eqref{eqn_multiview_complete_data_ll}. It is evident from this expression that, whenever a joint cluster has nonzero expected membership, the objective diverges to negative infinity if any of the entries \( \pi_{k^{(1)}, \ldots, k^{(V)}} \) approach zero. This implies that \( \log(\pi_{k^{(1)}, \ldots, k^{(V)}}) \) acts as a barrier-like term that discourages cluster probabilities from collapsing to zero during optimization.. Therefore, to encourage sparsity in the tensor \( \pi \), it is necessary to introduce a negative log-penalty that counteracts the effect of the log-barrier term. \cite{Huang2017ModelSelectionGMM} proposed using a penalty of the form \( \log(\delta + \cdot) \), where \( \delta > 0 \) is a small constant, in the context of single-view mixture models. This approach was later extended to the multi-view mixture model setting by \cite{carmichael2020learning}. More specifically, rather than maximizing the standard log-likelihood in \eqref{eqn_log_likelihod_multiview}, we instead optimize a penalized version that incorporates a sparsity-inducing term. The penalized log-likelihood takes the form:
\begin{equation}
     \ell(\{x_i\}_{i=1}^{N} \mid \boldsymbol{\varphi}, \pi) - \lambda \left( \sum_{k^{(1)}=1}^{K^{(1)}} \cdots \sum_{k^{(V)}=1}^{K^{(V)}} \log \left(  \delta +  \pi_{k^{(1)}, \ldots, k^{(V)}} \right) \right),
\end{equation}
where \( \lambda > 0 \) controls the strength of the sparsity penalty. The term $- \log \left(  \delta +  \pi_{k^{(1)}, \ldots, k^{(V)}} \right) $  has the effect of shrinking small values of \( \pi_{k^{(1)}, \dots, k^{(V)}} \) with sufficiently small components toward zero. Theoretical results in \cite{carmichael2020learning} demonstrate that, for sufficiently small values of $\delta$, the optimization of the penalized log-likelihood is insensitive to the specific choice of $\delta$.

This optimization problem can be addressed using an EM algorithm similar to that used for the unpenalized log-likelihood in \eqref{eqn_log_likelihod_multiview} and is derived in \cite{carmichael2020learning}. Specifically, the E-step remains unchanged from the unpenalized case, and the M-step for updating the cluster parameters \( \boldsymbol{\varphi} \) is identical to the update given in Eq. \eqref{eqn_multiview_m_step_clus_para}. The only modification to the EM algorithm lies in the update of \( \pi \) during the M-step. Given the expected cluster assignment probabilities \( \gamma^s(k^{(1)}, \ldots, k^{(V)} \mid x_i) \) for each $x_i$, as computed in the E-step (see Eq. \eqref{eqn_multiview_cluster_assign_prob}), the update for \( \pi \) is given by
\begin{equation}
\label{eqn_multiview_pi_update_pen}
    \pi^{s+1}_{k^{(1)},\ldots,k^{(V)}} = \frac{\left( \frac{1}{N} \sum_{i=1}^{N} \gamma^s(k^{(1)}, \ldots, k^{(V)} \mid x_i) - \lambda \right)_{+}}{ \sum_{j^{(1)}=1}^{K^{(1)}} \cdots \sum_{j^{(V)}=1}^{K^{(V)}} \left( \frac{1}{N} \sum_{i=1}^{N} \gamma^s(j^{(1)}, \ldots, j^{(V)} \mid x_i) - \lambda \right)_{+} },
\end{equation}
where \( (\cdot)_{+} \) denotes the positive part function, i.e., \( (x)_{+} = \max(x, 0) \). To ensure that the denominator in Eq. \eqref{eqn_multiview_pi_update_pen} is strictly positive, it suffices to choose the regularization parameter \( \lambda \) such that
\[
0 < \lambda < \frac{1}{\prod_{v=1}^{V} K^{(v)}},
\]
which ensures that at least one joint component retains positive mass after thresholding.

\section{Methodology}
\label{sec_method}
\subsection{Model Specification}
Section~\ref{sec_multiview_mixture} presents the general framework of the multi-view mixture model. Here, we focus on a special case with two views. In contrast to most existing work on multi-view clustering, which typically assumes that all views consist of fixed-dimensional feature vectors, our setting combines a fixed-dimensional view with a longitudinal view. We consider a dataset consisting of observations $\{x_i\}_{i=1}^{N}$, where each observation is of the form $x_i = (x_i^{(1)}, x_i^{(2)})$. Here, $x_i^{(1)} \in \mathbb{R}^{d^{(1)}}$ denotes a fixed-dimensional feature vector, and $x_i^{(2)} = \{x_{i,j}^{(2)}\}_{j=1}^{n_i}$ represents longitudinal measurements recorded at individual-specific time points $\{t_{i,j}\}_{j=1}^{n_i}$. Let \( \{(\xi_i^{(1)}, \xi_i^{(2)})\}_{i=1}^{N} \) denote the corresponding latent cluster memberships for all $N$ observations.

To instantiate the multi-view mixture model in this setting, we specify the distributional form for each view. For the fixed-dimensional view, we assume that observations within each cluster follow a multivariate normal distribution. Specifically, conditional on mixture component label $\xi_i^{(1)} = k$, we assume  
\begin{equation}
\label{eqn_model_fixed_dim}
x_i^{(1)} \sim \phi_{\text{MVN}}(\cdot \mid \mu_k, \Sigma_k)     
\end{equation}
 where \( \phi_{\text{MVN}}(\cdot \mid \mu_k, \Sigma_k) \) denotes the density of a multivariate normal distribution with mean vector \( \mu_k \) and covariance matrix \( \Sigma_k \).

Although the assumption of normality for each mixture component may be restrictive, particularly since the fixed dimensional features in our application include both numerical and categorical variables that do not naturally conform to this assumption, we mitigate this limitation using the PCAmix approach \cite{Chavent2014PCAmixdata}. PCAmix combines Principal Component Analysis (PCA) for numerical variables with Multiple Correspondence Analysis (MCA) for categorical variables within a unified generalized singular value decomposition framework. This approach produces a low-dimensional continuous representation of the features that captures the key variance structure while reducing noise and redundancy. These transformed features are then used in the multi-view mixture model.

For the view comprising longitudinal observations, we model the data within each cluster using a neural ODE framework as introduced in Section~\ref{sec_neural_ode}. Specifically, for each mixture component, we define a cluster-specific latent mean trajectory \( z_k: [0, T] \rightarrow \mathbb{R} \) governed by a neural ODE of the form 
\begin{equation}
\label{eqn_model_neural_ode}
\frac{dz_k(t)}{dt} = f_{\theta_k}(z_k(t), t),    
\end{equation}
with initial condition 
\begin{equation}
\label{eqn_model_initial_cond}
 z_{k,0} := z_k(0) ,
\end{equation}
where \( f_{\theta_k} \) is a neural network parameterized by cluster-specific parameters \( \theta_k \), as described in equation~\eqref{eqn_neural_ode_diff_eq}. We adopt a feedforward neural network architecture for \( f_{\theta_k} \), consisting of two hidden layers, as specified in \eqref{eqn_feedforward_nn}. The parameters \( \theta_k \) therefore include the weights and biases of the network. The latent trajectory \( z_k(t) \) is obtained by numerically solving the differential equation using the Runge--Kutta method \eqref{eqn_nerual_ode_solver}.

Now, conditional on the latent cluster label \( \xi_i^{(2)} = k \), we assume that a longitudinal observation \( x_{i,j}^{(2)} \) recorded at time \( t_{i,j} \) follows a normal distribution with mean given by the cluster-specific latent trajectory \( z_k(t_{i,j}) \) and variance \( \sigma_k^2 \), as given in Eq.~\eqref{eqn_normal_dist_longitudinal}:
\begin{equation}
\label{eqn_model_longitudinal}
x_{i,j}^{(2)} \sim \phi_N(x_{i,j}^{(2)} \mid z_k(t_{i,j}), \sigma_k^2).
\end{equation}
Given that we have specified the number of mixture components \( K^{(1)} \) and \( K^{(2)} \) for the two views, respectively, we define the set of mixture component parameters as
\[
\boldsymbol{\varphi} = \left\{ \{ \mu_k \}_{k=1}^{K^{(1)}}, \{ \Sigma_k \}_{k=1}^{K^{(1)}}, \{ \theta_k \}_{k=1}^{K^{(2)}}, \{z_{k,0}\}_{k=1}^{K^{(2)}}, \{ \sigma_k \}_{k=1}^{K^{(2)}} \right\},
\]
and let \( \pi \in \mathbb{R}^{K^{(1)} \times K^{(2)}} \) denote the joint distribution over view-specific cluster labels, as defined in Eq.~\eqref{eqn_cluster_prob_tensor}. The log-likelihood function is thus given by
\begin{equation}
\label{eqn_loglik_model}
    \ell(\{x_i\}_{i=1}^{N} | \boldsymbol{\varphi}, \pi) = \sum_{i=1}^{N} \log \left( \sum_{k^{(1)}=1}^{K^{(1)}} \sum_{k^{(2)}=1}^{K^{(2)}} \pi_{k^{(1)},k^{(2)}} \phi_{\text{MVN}}(x_{i}^{(1)}|\mu_{k^{(1)}}, \Sigma_{k^{(1)}}) \prod_{j=1}^{n_i} \phi_N(x_{i,j}^{(2)}|z_{k^{(2)}}(t_{i,j}), \sigma_{k^{(2)}}^2)\right) .
\end{equation}
Here, the dependence of the latent trajectory \( z_{k^{(2)}} \) on the neural ODE parameter \( \theta_{k^{(2)}} \) and initial condition $z_{k^{(2)},0}$ is implicit.

\subsection{EM Algorithm}
We adapt the EM algorithm described in Section~\ref{sec_background_em} and~\ref{sec_background_sparsity} for the general multi-view mixture model to our specific setting, where one view consists of fixed-dimensional features and the other comprises longitudinal measurements. We first consider maximizing the log-likelihood in \eqref{eqn_loglik_model}, without incorporating any sparsity-inducing penalty. Conditional on current parameter estimate $(\boldsymbol{\varphi}^s, \pi^s)$, the expected complete data log-likelihood takes the form
\begin{equation}
    Q(\boldsymbol{\varphi}, \pi|\boldsymbol{\varphi}^s, \pi^s)   = \sum_{i=1}^{N} \sum_{k^{(1)}=1}^{K^{(1)}} \sum_{k^{(2)}=1}^{K^{(2)}} \gamma^s(k^{(1)},k^{(2)}|x_i) \log \left( \pi_{k^{(1)},k^{(2)}} \phi_{\text{MVN}}\left( x_i^{(1)}|\mu_{k^{(1)}}, \Sigma_{k^{(1)}} \right) \prod_{j=1}^{n_i} \phi_N \left( x_{i,j}^{(2)} | z_{k^{(2)}}(t_{i,j}), \sigma_{k^{(2)}}^2 \right)  \right) ,
\end{equation}
where 
\begin{equation}
\label{eqn_e_step_model}
\gamma^s(k^{(1)},k^{(2)}|x_i) = \frac{\pi^s_{k^{(1)},k^{(2)}} \phi_{\text{MVN}}\left( x_i^{(1)}|\mu^s_{k^{(1)}}, \Sigma^s_{k^{(1)}} \right) \prod_{j=1}^{n_i} \phi_N \left( x_{i,j}^{(2)} | z^s_{k^{(2)}}(t_{i,j}), (\sigma^s_{k^{(2)}})^2 \right)}{\sum_{l^{(1)}=1}^{K^{(1)}} \sum_{l^{(2)}=1}^{K^{(2)}} \pi^s_{l^{(1)},l^{(2)}} \phi_{\text{MVN}}\left( x_i^{(1)}|\mu^s_{l^{(1)}}, \Sigma^s_{l^{(1)}} \right) \prod_{j=1}^{n_i} \phi_N \left( x_{i,j}^{(2)} | z^s_{l^{(2)}}(t_{i,j}), (\sigma^s_{l^{(2)}})^2 \right) }
\end{equation}
is the probability that observation $x_i$ belongs to cluster $k^{(1)}$ in view 1 and cluster $k^{(2)}$ in view 2 under the current parameter estimates \( (\boldsymbol{\varphi}^s, \pi^s) \). 

In the M-step, we maximize the expected complete-data log-likelihood $Q(\boldsymbol{\varphi}, \pi \mid \boldsymbol{\varphi}^s, \pi^s)$ with respect to the parameters $\boldsymbol{\varphi}$ and $\pi$. As discussed in Section~\ref{sec_background_em}, this optimization problem is separable, allowing us to update the cluster parameters for each view and the joint cluster probability tensor $\pi$ independently.

The update rule for $\pi$ is given by:
\begin{equation}
\label{eqn_update_pi}
    \pi_{k^{(1)}, k^{(2)}}^{s+1} = \frac{1}{N} \sum_{i=1}^{N} \gamma^s(k^{(1)}, k^{(2)} \mid x_i).
\end{equation}
For each cluster $k^{(1)} = 1, \ldots, K^{(1)}$ in the first view, the maximization of the expected complete-data log-likelihood $Q(\boldsymbol{\varphi}, \pi \mid \boldsymbol{\varphi}^s, \pi^s)$ with respect to the mean $\mu_{k^{(1)}}$ and covariance $\Sigma_{k^{(1)}}$ admits closed-form solutions, given by:
\begin{equation}
\label{eqn_m_step_mu_static}
\mu_{k^{(1)}}^{s+1} = \frac{ \sum_{i=1}^{N} \sum_{k^{(2)}=1}^{K^{(2)}} \gamma^{s}(k^{(1)}, k^{(2)} \mid x_i) \, x_i^{(1)} }{ \sum_{i=1}^{N} \sum_{k^{(2)}=1}^{K^{(2)}} \gamma^{s}(k^{(1)}, k^{(2)} \mid x_i)},
\end{equation}
\begin{equation}
\label{eqn_m_step_sigma_static}
\Sigma_{k^{(1)}}^{s+1} = \frac{ \sum_{i=1}^{N} \sum_{k^{(2)}=1}^{K^{(2)}} \gamma^{s}(k^{(1)}, k^{(2)} \mid x_i) \, (x_i^{(1)} - \mu_{k^{(1)}}^{s+1}) (x_i^{(1)} - \mu_{k^{(1)}}^{s+1})^{T} }{ \sum_{i=1}^{N} \sum_{k^{(2)}=1}^{K^{(2)}} \gamma^{s}(k^{(1)}, k^{(2)} \mid x_i) }.
\end{equation}

These update formulas closely resemble those of the classical EM algorithm for single-view Gaussian mixture models.

Next, we consider the second view, which consists of longitudinal observations. For any mixture component $k^{(2)} = 1, \ldots, K^{(2)}$, our goal is to maximize the expected complete-data log-likelihood $Q(\boldsymbol{\varphi}, \pi \mid \boldsymbol{\varphi}^s, \pi^s)$ with respect to the neural network parameters $\theta_{k^{(2)}}$, the initial condition $z_{k^{(2)},0}$, and variance $\sigma_{k^{(2)}}^2$ associated with component $k^{(2)}$. The contribution of the $k^{(2)}$th longitudinal component to $Q(\boldsymbol{\varphi}, \pi \mid \boldsymbol{\varphi}^s, \pi^s)$ can be written as

\begin{eqnarray}
 Q(\boldsymbol{\varphi}, \pi|\boldsymbol{\varphi}^s, \pi^s)   &=& \sum_{i=1}^{N} \sum_{k^{(1)}=1}^{K^{(1)}} \gamma^s(k^{(1)},k^{(2)}|x_i)   \sum_{j=1}^{n_i} \log \left( \phi_N \left( x_{i,j}^{(2)} | z_{k^{(2)}}(t_{i,j}), \sigma_{k^{(2)}}^2 \right)  \right) + \text{Const}, \nonumber \\
 &=& \sum_{i=1}^{N} \sum_{k^{(1)}=1}^{K^{(1)}} \gamma^s(k^{(1)},k^{(2)}|x_i)   \sum_{j=1}^{n_i} \left( - \frac{ \left( x_{i,j}^{(2)} - z_{k^{(2)}}(t_{i,j}) \right)^2 }{  2 \sigma^2_{k^{(2)}}} - \frac{1}{2} \log \left(\sigma^2_{k^{(2)}} \right) \right)   + \text{Const},
\end{eqnarray}
where all terms not depending on the parameters of the $k^{(2)}$-th longitudinal component have been absorbed into the constant term.

We update the parameters $\sigma^2_{k^{(2)}}, z_{k^{(2)},0}, \theta_{k^{(2)}}$ iteratively by optimizing each while holding the others fixed. To begin, consider updating $\sigma^2_{k^{(2)}}$. Differentiating $Q(\boldsymbol{\varphi}, \pi \mid \boldsymbol{\varphi}^s, \pi^s)$ with respect to $\sigma^2_{k^{(2)}}$ and setting the derivative to zero yields the following closed-form solution:
\begin{equation}    
\label{eqn_m_step_sigma_long}
\sigma^2_{k^{(2)}} = \frac{ \sum_{i=1}^{N} \sum_{k^{(1)}=1}^{K^{(1)}} \gamma^s(k^{(1)}, k^{(2)} \mid x_i) \sum_{j=1}^{n_i} \left( x_{i,j}^{(2)} - z_{k^{(2)}}(t_{i,j}) \right)^2 }{ \sum_{i=1}^{N} n_i \sum_{k^{(1)}=1}^{K^{(1)}} \gamma^s(k^{(1)}, k^{(2)} \mid x_i) }.
\end{equation}
We update the neural network parameters $\theta_{k^{(2)}}$ while holding $z_{k^{(2)},0}$ and $\sigma^2_{k^{(2)}}$ fixed. Since a closed-form solution is not available for this update, we perform numerical optimization of $Q(\boldsymbol{\varphi}, \pi \mid \boldsymbol{\varphi}^s, \pi^s)$ using the Adam optimizer \citep{kingma2015adam}. The optimization procedure is terminated when the change in the associated loss between two consecutive iterations falls below a pre-specified threshold. Similarly, the update of the initial condition $z_{k^{(2)},0}$ is carried out by holding $\theta_{k^{(2)}}$ and $\sigma^2_{k^{(2)}}$ fixed and optimizing numerically using the Adam optimizer.
\\\\
\textbf{Sparsity Inducing Log Penalty}\\
Incorporating the sparsity-inducing log penalty discussed in Section~\ref{sec_background_sparsity}, we augment the log-likelihood in \eqref{eqn_loglik_model} with this penalty term to obtain the penalized log-likelihood:

\begin{equation}
   \label{eqn_pen_ll_model}
  \ell(\{x_i\}_{i=1}^{N} \mid \boldsymbol{\varphi}, \pi) - \lambda \left( \sum_{k^{(1)}=1}^{K^{(1)}} \sum_{k^{(2)}=1}^{K^{(2)}} \log \left(  \delta +  \pi_{k^{(1)}, k^{(2)}} \right) \right),
\end{equation}
for some tuning parameter $\lambda > 0$. Recall from Section \ref{sec_background_sparsity} that the E-step and the updates for $\boldsymbol{\varphi}$ in the M-step are identical to the case without log penalty. The update rule for $\pi$ in Equation \eqref{eqn_update_pi} is now replaced by the following expression:

\begin{equation}
\label{eqn_model_pi_update_pen}
    \pi^{s+1}_{k^{(1)},k^{(2)}} = \frac{\left( \frac{1}{N} \sum_{i=1}^{N} \gamma^s(k^{(1)}, k^{(2)} \mid x_i) - \lambda \right)_{+}}{ \sum_{j^{(1)}=1}^{K^{(1)}} \sum_{j^{(2)}=1}^{K^{(2)}} \left( \frac{1}{N} \sum_{i=1}^{N} \gamma^s(j^{(1)}, j^{(2)} \mid x_i) - \lambda \right)_{+} }.
\end{equation}
To ensure that the denominator remains positive, it is sufficient to choose $\lambda$ such that $0 < \lambda < \frac{1}{K^{(1)} K^{(2)}}.$

\begin{figure}[H]
\centering
\begin{minipage}{0.95\linewidth}
\caption{EM Algorithm for the Two-View Mixture Model with Sparsity-Inducing Log Penalty}
\label{alg_em_model_pen}

\textbf{Input:} Dataset $\{x_i\}_{i=1}^N$; convergence threshold $\epsilon > 0$

\vspace{0.5em}
\textbf{Initialize:} Cluster parameters $\boldsymbol{\varphi}$ and joint cluster probabilities $\pi$

\vspace{0.5em}
\textbf{Repeat}
\begin{enumerate}
    \item \textbf{E-step:} For each observation $x_i$, and for all $k^{(1)} = 1,\ldots,K^{(1)}$ and $k^{(2)} = 1,\ldots,K^{(2)}$, compute cluster assignment probabilities
    \[
        \gamma(k^{(1)}, k^{(2)} \mid x_i)
        \quad \text{using Equation \eqref{eqn_e_step_model}.}
    \]

    \item \textbf{M-step:}
    \begin{enumerate}
        \item For each $k^{(1)} = 1,\ldots,K^{(1)}$, update parameters $\mu_{k^{(1)}}$ and $\Sigma_{k^{(1)}}$ using Equations \eqref{eqn_m_step_mu_static} and \eqref{eqn_m_step_sigma_static}.

        \item For each $k^{(2)} = 1,\ldots,K^{(2)}$, update $\sigma_{k^{(2)}}^2$ using Equation \eqref{eqn_m_step_sigma_long}.

        \item Update $\theta_{k^{(2)}}$ and $z_{k^{(2)},0}$ via numerical optimization with the Adam optimizer.

        \item Update joint cluster probabilities $\pi_{k^{(1)}, k^{(2)}}$ for all $k^{(1)}, k^{(2)}$ using the penalized update rule in Equation \eqref{eqn_model_pi_update_pen}.

        \item Compute the penalized log-likelihood in Equation \eqref{eqn_pen_ll_model}.
    \end{enumerate}
\end{enumerate}

\textbf{Until:} Absolute difference in penalized log-likelihood between iterations is $< \epsilon$

\vspace{0.5em}
\textbf{Output:} Final estimates of parameters $\boldsymbol{\varphi}$, $\pi$, and cluster assignment probabilities $\gamma(k^{(1)}, k^{(2)} \mid x_i)$ for all $i, k^{(1)}, k^{(2)}$
\end{minipage}
\end{figure}
The EM algorithm for maximizing the penalized log-likelihood in Eq. \eqref{eqn_pen_ll_model} is summarized in Algorithm~\ref{alg_em_model_pen}. It begins by initializing the model parameters $\boldsymbol{\varphi}$ and $\pi$, which involves random sampling. The initializations for $\{ \mu_{k^{(1)}}, \Sigma_{k^{(1)}} \}_{k^{(1)}=1}^{K^{(1)}}, \{\sigma_{k^{(2)}}\}_{k^{(2)}=1}^{K^{(2)}},$ and $\{z_{k^{(2)},0}\}_{k^{(2)}=1}^{K^{(2)}}$ are relatively straightforward. However, in experiments we observe that the convergence behavior of the EM algorithm is more sensitive to the initialization of the neural network parameters $\{\theta_{k^{(2)}}\}_{k^{(2)}=1}^{K^{(2)}}$, which can lead to slow convergence in some runs. To mitigate this issue, we generate multiple random initializations of model parameters and run the EM algorithm for each. The solution corresponding to the highest penalized log-likelihood is then selected.

\subsection{Model Selection}
Selecting the optimal number of clusters is a fundamental aspect of mixture modeling, as statistical inference and subsequent interpretations can be highly sensitive to this decision. Standard model selection criteria, such as the Akaike Information Criterion (AIC) and Bayesian Information Criterion (BIC), are frequently used for this purpose~\cite{McLachlan2000, Celeux1992CEM, Biernacki2000}. However, these criteria may be unsuitable in our setting due to the use of neural networks for modeling longitudinal data. Both AIC and BIC apply penalties based on the number of model parameters, but neural networks typically involve a large number of parameters, many of which do not contribute equally to model complexity. As a result, the notion of ``effective model complexity'' is poorly captured by raw parameter counts, leading AIC and BIC to impose overly conservative penalties that may misrepresent the model's actual complexity.

Instead, we employ $K$-fold cross-validated log-likelihood as the model selection criterion for determining the number of mixture components. For each fold, models with different numbers of components are fitted on the training set, and their predictive performance is evaluated by computing the log-likelihood on the validation set. The configuration yielding the highest validation (penalized) log-likelihood is selected as the optimal model.

\section{Simulation Studies}
\label{sec_simulation}

We conduct two simulation studies to evaluate the proposed EM algorithm's ability to recover true model parameters and the effectiveness of cross-validated log-likelihood in identifying the correct number of mixture components. Additionally, a sensitivity analysis examines the impact of the sparsity tuning parameter \(\lambda\) on parameter estimates.

\subsection{Simulation Design}
\label{subsec:sim_design}
\paragraph{Simulation 1}
For the first simulation, we generate data from a two-view mixture model with two mixture components in each view. 
The two longitudinal components are defined by component-specific trajectory functions 
on $t \in [0,100]$:
\[
f_1(t)=\sin(t/50), \qquad f_2(t)=\cos(t/50)-1,
\]
with Gaussian measurement noise $\epsilon(t)\sim \mathcal{N}(0,\sigma_k^2)$, 
where $\sigma_k^2=0.1$ for both clusters. For each individual, the number of longitudinal measurements is drawn from the discrete uniform distribution $\mathcal{U}_d(6, 20)$, and the measurement times are sampled from the continuous uniform distribution $\mathcal{U}(0, 100)$.

The two fixed-dimensional mixture components are bivariate Gaussian distributions with parameters
\[
\mu_1 = \begin{pmatrix} 10 \\ 0 \end{pmatrix}, \quad 
\mu_2 = \begin{pmatrix} 12 \\ 2 \end{pmatrix}, \quad
\Sigma_1 = \Sigma_2 = I_2,
\]
where \(I_2\) is the \(2 \times 2\) identity matrix. The joint cluster membership probability matrix is
\[
\pi = \mathrm{diag}\left(\frac{1}{2}, \frac{1}{2}\right),
\]
with \(\mathrm{diag}(\cdot)\) denoting a diagonal matrix.

\paragraph{Simulation 2}
For the second simulation, the longitudinal trajectories are
\[
f_1(t)=\sin(t/50), \qquad 
f_2(t)=\cos(t/50)-1, \qquad 
f_3(t)=0.5,
\]
with Gaussian noise $\epsilon(t)\sim \mathcal{N}(0,\sigma_k^2)$ and $\sigma_k^2=0.1$ for both clusters. As in Simulation 1, each individual has a number of longitudinal measurements drawn from \(\mathcal{U}_d(6, 20)\), with measurement times sampled from \(\mathcal{U}(0, 100)\).

The three fixed-dimensional mixture components are bivariate Gaussian distributions with parameters
\[
\mu_1 = \begin{pmatrix} 10 \\ 0 \end{pmatrix}, \quad
\mu_2 = \begin{pmatrix} 12 \\ 2 \end{pmatrix}, \quad
\mu_3 = \begin{pmatrix} 8 \\ -2 \end{pmatrix}, \qquad
\Sigma_1 = \Sigma_2 = \Sigma_3 = I_2,
\]
where \(I_2\) is the \(2 \times 2\) identity matrix. The joint cluster membership probability matrix is
\[
\pi = \mathrm{diag}\left(\frac{1}{3}, \frac{1}{3}, \frac{1}{3}\right).
\]

\subsection{Simulation Results}

\subsubsection{Selection of Number of Mixture Components}
\label{subsec:cluster_selection}
For each simulation setting, we generate datasets with sample size $N=1000$, and fit models with varying numbers of mixture 
components $(K^{(1)} \times K^{(2)})$ using the EM algorithm with sparsity-inducing log penalty. Model selection 
is performed using five-fold cross-validated log-likelihood, and the 
configuration $(K^{(1)} \times K^{(2)})$ with the largest value is selected. The sparsity parameter $\lambda$ is fixed at $\lambda = 0.1$ and the sensitivity of the results to the choice of $\lambda$ is examined in Section~\ref{subsec:sparsity_sensitivity}. 
Tables~\ref{tab:sim1_results} and~\ref{tab:sim2_results} report the 
cross-validated log-likelihood values for Simulations~1 and~2, respectively. 
In both cases, the configuration with the highest cross-validated log-likelihood 
coincides with the true model, indicating that cross-validated log-likelihood 
provides a reasonable criterion for model selection.

\begin{table}[H]
\centering
\caption{Five-fold cross-validated log-likelihood for Simulation 1 with Sample size N = 1000.}
\label{tab:sim1_results}
\small
\begin{tabular}{lrr}
\toprule
Configuration $(K^{(1)}\times K^{(2)})$ 
& CV Log-likelihood 
& CV Log-lik / Subject \\
\midrule
1$\times$1 & -13240 & -66.20 \\
1$\times$2 & -14096 & -70.48 \\
1$\times$3 & -15880  & -79.40 \\

2$\times$1 & -12460 & -62.30 \\
\textbf{2$\times$2} & \textbf{-9299} & \textbf{-46.50} \\
2$\times$3 & -11335 & -56.68 \\

3$\times$1 & -15720 & -78.60 \\
3$\times$2 & -13318 & -66.59 \\
3$\times$3 & -10542 & -52.71 \\
\bottomrule
\end{tabular}
\end{table}

\begin{table}[H]
\centering
\caption{Five-fold cross-validated log-likelihood for Simulation 2 with Sample size N = 1000}
\label{tab:sim2_results}
\small
\begin{tabular}{lrr}
\toprule
Configuration $(K^{(1)}\times K^{(2)})$ 
& CV Log-likelihood 
& CV Log-lik / Subject \\
\midrule
1$\times$1 & -18994 & -94.97 \\
1$\times$2 & -17286 & -86.43 \\
1$\times$3 & -16240 & -81.20 \\
1$\times$4 & -15739 & -78.70 \\

2$\times$1 & -15740 & -78.70 \\
2$\times$2 & -11061 & -55.30 \\
2$\times$3 & -12029 & -60.15 \\
2$\times$4 & -11226 & -56.13\\

3$\times$1 & -15880 & -79.40 \\
3$\times$2 & -11926 & -59.63 \\
\textbf{3$\times$3} & \textbf{-8392} & \textbf{-41.90} \\
3$\times$4 & -10082 & -50.41 \\

4$\times$1 & -15032 & -75.16 \\
4$\times$2 & -10904 & -54.52 \\
4$\times$3 & -9228 & -46.14 \\
4$\times$4 & -9983 & -49.92\\

\bottomrule
\end{tabular}
\end{table}

\subsubsection{Parameter Recovery}
\label{subsec:param_recovery}

Next, we evaluate the accuracy of parameter estimation under the correctly 
specified cluster configurations, examining whether the EM algorithm recovers 
the true model parameters. To assess the effect of sample size on estimation 
accuracy, we generate datasets with sample sizes 
$N \in \{50,100,300,500,1000\}$; for each $N$, five replicates of datasets are simulated. Estimation accuracy is measured by the distance between the estimated 
parameters $(\hat{\mu}_k, \hat{\Sigma}_k, \hat{\pi}, \hat{f}_l)$ and the 
corresponding true parameters $(\mu_k, \Sigma_k, \pi, f_l)$, where 
$k,l \in \{1,2\}$ in Simulation~1 and $k,l \in \{1,2,3\}$ in Simulation~2. 
For $\mu_k$, we use the Euclidean norm $\|\cdot\|_2$; for $\Sigma_k$ and $\pi$, 
the Frobenius norm $\|\cdot\|_F$; and for the trajectory $f_l$, the 
$L_2([0,T])$ distance computed over a discretized grid. Furthermore, we assess clustering accuracy using the adjusted Rand index 
(ARI) \citep{hubert1985comparing}.

\begin{table}[H]
\centering
\caption{Simulation Setting 1 (2x2 setting, $\lambda = 0.1$). Results are averaged over 5 replicates experiments with standard deviations in parentheses.}
\label{tab:sim_2x2_N}
\small
\setlength{\tabcolsep}{6pt}
\begin{tabular}{lccccc}
\toprule
$N$ &
$\sum_k \|\hat{\mu}_k-\mu_k\|_2$ &
$\sum_k\|\hat{\Sigma}_k-\Sigma_k\|_F$ &
$\|\hat{\pi}-\pi\|_F$ &
$\sum_l \|\hat{f}_l-f\|_{L_2(0,T)}$ &
ARI \\
\midrule
50  & 0.492 (0.141) & 0.603 (0.176) & 0.023 (0.008) & 0.214 (0.071) & 1.000 (0.000) \\
100 & 0.334 (0.096) & 0.412 (0.121) & 0.008 (0.005) & 0.151 (0.049) & 1.000 (0.000) \\
300 & 0.221 (0.066) & 0.281 (0.083) & 0.006 (0.004) & 0.103 (0.032) & 1.000 (0.000) \\
500 & 0.171 (0.052) & 0.216 (0.067) & 0.001 (0.000) & 0.081 (0.025) & 1.000 (0.000) \\
1000 & 0.122 (0.038) & 0.161 (0.052) & 0.000 (0.000) & 0.067 (0.020) & 1.000 (0.000) \\
\bottomrule

\end{tabular}
\end{table}

\begin{table}[H]
\centering
\caption{Simulation Setting 2 (3x3 setting, $\lambda = 0.1$). Results are averaged over 5 replicates experiments with standard deviations in parentheses.}
\label{tab:sim_3x3_N}
\small
\setlength{\tabcolsep}{6pt}
\begin{tabular}{lccccc}
\toprule
$N$ &
$\sum_k \|\hat{\mu}_k-\mu_k\|_2$ &
$\sum_k\|\hat{\Sigma}_k-\Sigma_k\|_F$ &
$\|\hat{\pi}-\pi\|_F$ &
$\sum_l \|\hat{f}_l-f\|_{L_2(0,T)}$ &
ARI \\
\midrule
50  & 0.781 (0.213) & 0.944 (0.258) & 0.013 (0.006) & 0.322 (0.098) & 1.000 (0.000) \\
100 & 0.552 (0.154) & 0.684 (0.192) & 0.008 (0.002) & 0.244 (0.074) & 1.000 (0.000) \\
300 & 0.381 (0.108) & 0.486 (0.139) & 0.003 (0.002) & 0.181 (0.055) & 1.000 (0.000) \\
500 & 0.296 (0.086) & 0.382 (0.111) & 0.003 (0.001) & 0.151 (0.044) & 1.000 (0.000) \\
1000 & 0.214 (0.064) & 0.281 (0.084) & 0.001 (0.000) & 0.114 (0.033) & 1.000 (0.000) \\
\bottomrule

\bottomrule
\end{tabular}
\end{table}
Tables~\ref{tab:sim_2x2_N} and~\ref{tab:sim_3x3_N} summarize the estimation results  for Simulation~1 and~2, respectively, across different sample sizes $N$. In both settings, estimation accuracy improves consistently for all model parameters, 
with the distances approaching zero as the sample size increases. This demonstrates 
that the EM algorithm can recover the true parameters given a sufficiently large sample.  Furthermore, in both settings, the ARI equals 1, indicating perfect recovery of the 
clustering structure.

\subsubsection{Sensitivity to the Sparsity Tuning Parameter}
\label{subsec:sparsity_sensitivity}
Finally, we investigate the sensitivity of parameter estimation to the sparsity 
tuning parameter $\lambda$. In particular, we assess the accuracy of the estimated parameters for various values of $\lambda$. 
\begin{table}[H]
\centering
\caption{Sensitivity of estimation and clustering performance to the sparsity tuning parameter $\lambda$ in Simulation Setting 1 with Sample size N fixed at 1000 ($2\times2$ model).Results are reported as mean (standard deviation) over simulation replicates.}
\label{tab:lambda_2x2}
\small
\setlength{\tabcolsep}{6pt}
\begin{tabular}{lccccc}
\toprule
$\lambda$  &
$\sum_k \|\hat{\mu}_k-\mu_k\|_2$ &
$\sum_k\|\hat{\Sigma}_k-\Sigma_k\|_F$ &
$\|\hat{\pi}-\pi\|_F$ &
$\sum_l \|\hat{f}_l-f\|_{L_2(0,T)}$ &
ARI \\
\midrule
0.00  & 0.168 (0.052) & 0.214 (0.071) & 0.021 (0.008) & 0.089 (0.028) & 1.000 (0.000) \\
0.05  & 0.139 (0.043) & 0.178 (0.060) & 0.009 (0.004) & 0.074 (0.023) & 1.000 (0.000) \\
0.10  & 0.122 (0.038) & 0.161 (0.052) & 0.000 (0.000) & 0.067 (0.020) & 1.000 (0.000) \\
0.20  & 0.151 (0.047) & 0.195 (0.065) & 0.013 (0.006) & 0.081 (0.026) & 1.000 (0.000) \\
0.50  & 0.231 (0.071) & 0.298 (0.094) & 0.084 (0.021) & 0.124 (0.039) & 1.000 (0.000) \\
\bottomrule
\end{tabular}
\end{table}

\begin{table}[H]
\centering
\caption{Sensitivity of estimation and clustering performance to the sparsity tuning parameter $\lambda$ in Simulation Setting 2 ($3\times3$ model) with $N=1000$. Results are reported as mean (standard deviation) over simulation replicates.}
\label{tab:lambda_3x3}
\small
\setlength{\tabcolsep}{6pt}
\begin{tabular}{lcccccc}
\toprule
$\lambda$ &
$\sum_k \|\hat{\mu}_k-\mu_k\|_2$ &
$\sum_k\|\hat{\Sigma}_k-\Sigma_k\|_F$ &
$\|\hat{\pi}-\pi\|_F$ &
$\sum_l \|\hat{f}_l-f\|_{L_2(0,T)}$ &
ARI \\
\midrule
0.00 &  0.283 (0.082) & 0.372 (0.109) & 0.034 (0.011) & 0.151 (0.044) & 1.000 (0.000) \\
0.05  & 0.239 (0.071) & 0.318 (0.094) & 0.015 (0.006) & 0.128 (0.037) & 1.000 (0.000) \\
0.10  & 0.214 (0.064) & 0.281 (0.084) & 0.001 (0.000) & 0.114 (0.033) & 1.000 (0.000) \\
0.20   & 0.247 (0.073) & 0.333 (0.097) & 0.019 (0.008) & 0.137 (0.040) & 1.000 (0.000) \\
0.50   & 0.392 (0.114) & 0.528 (0.152) & 0.103 (0.029) & 0.201 (0.058) & 1.000 (0.000) \\
\bottomrule
\end{tabular}
\end{table}
Tables~\ref{tab:lambda_2x2} and~\ref{tab:lambda_3x3} report the estimation and clustering results for Simulation~1 and~2, respectively, with the sample size fixed 
at $N=1000$. Across both simulation settings, parameter estimation accuracy initially 
improves as $\lambda$ increases, reaching its optimum at $\lambda = 0.1$, after which 
accuracy declines.

\section{Data Application}
\label{sec_data_app}
The dataset analyzed in this study consists of longitudinal follow-up measurements and baseline clinical characteristics of patients diagnosed with antineutrophil cytoplasmic antibody (ANCA)–associated vasculitis (AAV) in Ireland\cite{Scott2022}. The patient cohort includes individuals diagnosed and managed at tertiary nephrology and vasculitis referral centers across the country from 2012 to 2026, with data harmonized from institutional electronic health records and local research registries. The primary longitudinal biomarker analyzed was serum creatinine, a routinely measured indicator of renal function that reflects both disease activity and renal recovery trajectories over time. The observation period for each patient ranged from 180 days to 3 years following diagnosis, encompassing the subacute post-induction phase as well as subsequent maintenance or relapse periods. The distribution of follow-up encounters and creatinine values is illustrated in Figure 1, where some patients exhibit densely recorded trajectories with near-monthly follow-up, while others display only a few irregular measurements over several years. Consequently, the longitudinal data are characterized by sparse and unevenly spaced observations, with substantial heterogeneity in both sampling frequency and temporal alignment across patients. This irregular sampling poses significant challenges for traditional trajectory-based clustering methods, which typically assume evenly spaced measurement times. In addition to the longitudinal measurements, we included 17 baseline static covariates as previously proposed \citep{Karl2023}. The complete list of these variables is provided in Appendix A. As described in Section 3.1, these baseline features—comprising both categorical and numerical variables—were processed using the PCAmix algorithm to account for the mixed data structure. Following the dimension-reduction procedure, the baseline data were projected onto a five-dimensional subspace. The number of retained components was determined by examining the eigenvalues and the cumulative proportion of explained variance. The first five components together accounted for approximately 72 \(\%\) of the total variability in the baseline feature space, beyond which the marginal gain from additional components was minimal. 

\begin{figure}[H]
    \centering
    \includegraphics[width=1.0\linewidth]{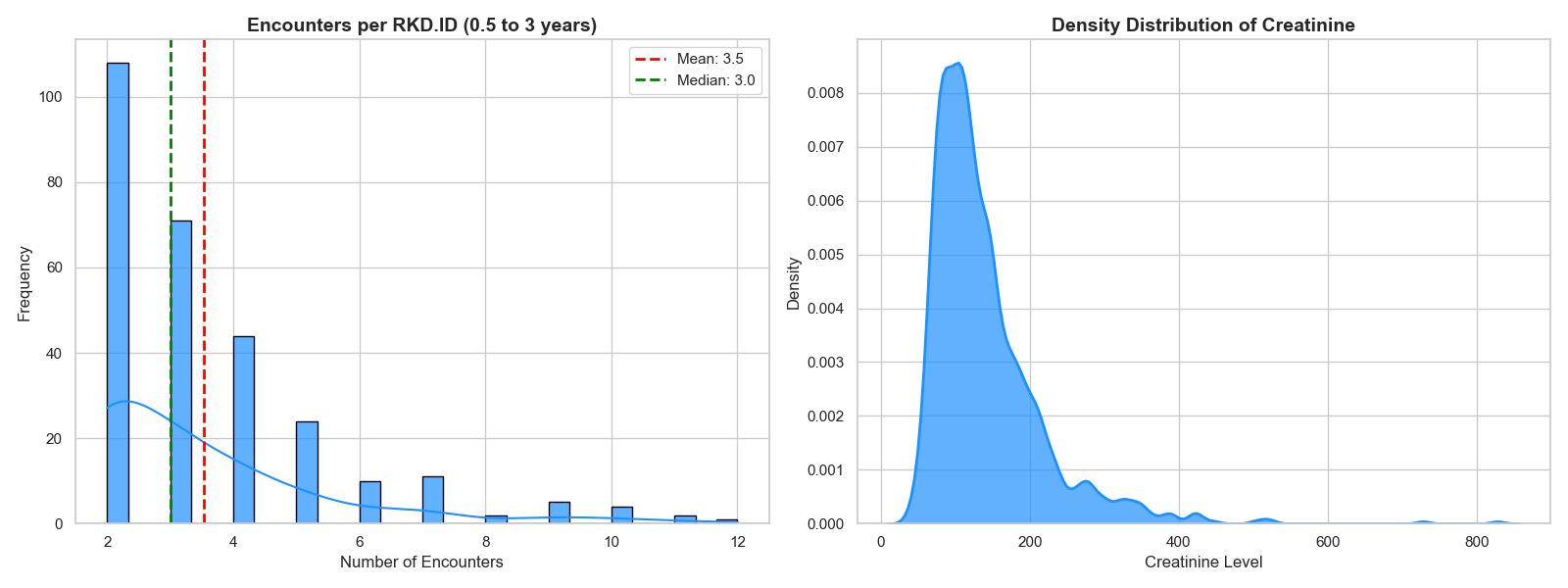}
    \caption{Explanatory Data Analysis}
    \label{fig:EDA}
\end{figure}

We apply the proposed multiview mixture model using the EM algorithm with a sparsity-inducing log penalty, setting \(\lambda = 0.1\), to this dataset. We evaluate a range of cluster configurations using five-fold cross-validated log-likelihood to determine the appropriate cluster structure for the creatinine data. As shown in Table~\ref{tab:sim1_results_real}, the \(2\times2\) configuration achieves the highest cross-validated log-likelihood. Increasing the number of longitudinal clusters beyond two does not improve predictive performance under cross-validation. We therefore adopt the \(2\times2\) configuration for the subsequent analyses.

\begin{table}[H]
\centering
\caption{Five-fold cross-validated log-likelihood for Real Data.}
\label{tab:sim1_results_real}
\small
\begin{tabular}{lrr}
\toprule
Configuration $(K^{(1)}\times K^{(2)})$ 
& CV Log-likelihood 
& CV Log-lik / Subject \\
\midrule
1$\times$1 &-13364  & -238.64 \\
1$\times$2 & -13478 &  -240.68\\
1$\times$3 & -15920  &  -284.29\\

2$\times$1 & -14247 & -254.41 \\
\textbf{2$\times$2} & \textbf{-12651} & \textbf{-225.91} \\
2$\times$3 &-13610  &  -243.12\\

3$\times$1 &-18052  & -322.36 \\
3$\times$2 & -17047 &  -304.41\\
3$\times$3 & -15651 & -273.52 \\
\bottomrule
\end{tabular}
\end{table}

\begin{figure}[H]
    \centering
    \includegraphics[width=1.0 \linewidth]{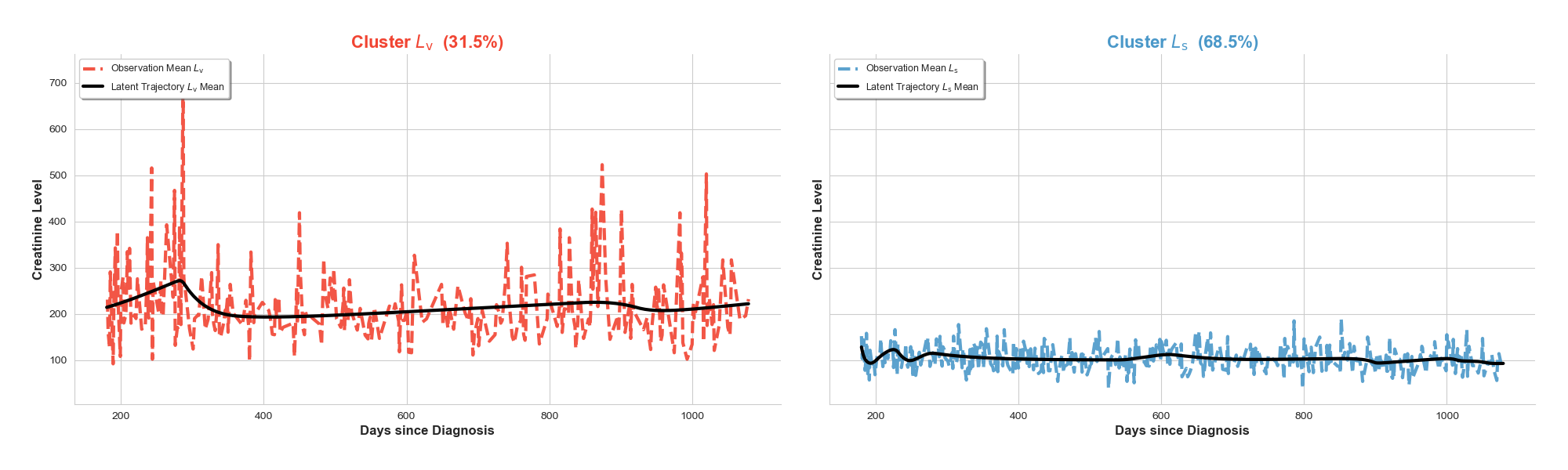}
    \caption{Latent Cluster for creatinine covering from 180 days to 3 years (2 clusters)}
    \label{fig:2x2_1080}
\end{figure}

 Figure~\ref{fig:2x2_1080} shows the estimated latent creatinine trajectories for the two longitudinal clusters, alongside the mean observed trajectories, calculated as the average creatinine level at each time point for patients within each cluster. The first longitudinal cluster (red in Figure~\ref{fig:2x2_1080}), comprising about \(30\%\) of patients, shows substantially higher and more variable creatinine levels over time, whereas the second cluster (blue) exhibits a stable profile with consistently lower creatinine values around \(100~\mu\mathrm{mol/L}\).

\begin{figure}[h]
    \centering
    \includegraphics[width= 1.0\linewidth]{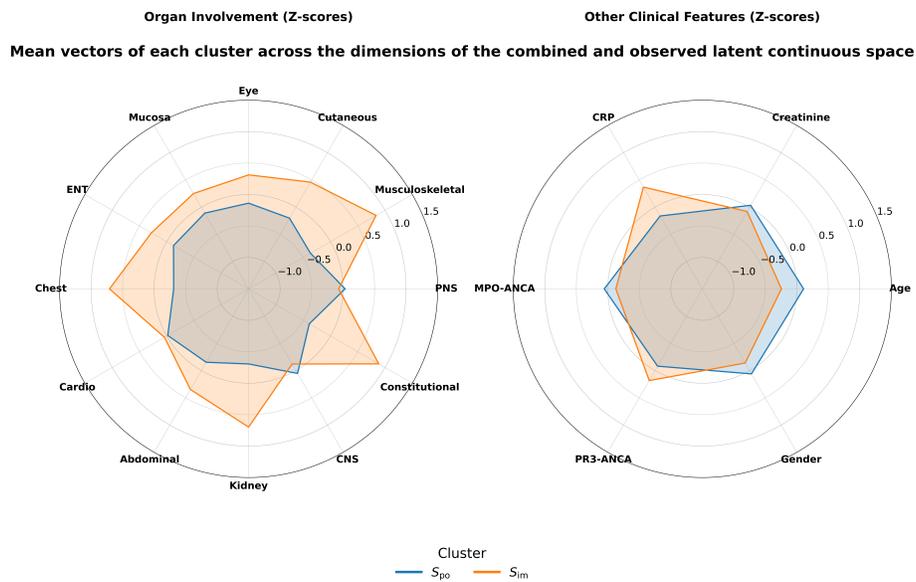}
    \caption{Mean latent representations of each cluster across the dimensions of the combined and observed continuous latent space under the 2-cluster configuration}
    \label{fig:2x2_static}
\end{figure}

Clear differences in baseline characteristics were observed between the two baseline clusters as shown in Figure~\ref{fig:2x2_static}. The first cluster, $S_{\mathrm{po}}$ (Static group with Pauci-Organ Low Inflammation, $n=196$), exhibited consistently lower rates of extra-renal involvement. Constitutional, musculoskeletal, and chest manifestations were present in only around 10--20\% of patients. In contrast, the second cluster, $S_{\mathrm{im}}$ (Static group with Inflammatory Multi-system, $n=86$), showed markedly higher frequencies of multi-system involvement, with most manifestations occurring in approximately 60--75\% of patients (see Table B2 in the Appendix for details). This pattern suggests a more systemic inflammatory phenotype.  Serologically, PR3-ANCA positivity was more common in $S_{\mathrm{im}}$, whereas MPO-ANCA predominated in $S_{\mathrm{po}}$. Serum CRP levels were also noticeably higher in $S_{\mathrm{im}}$, while differences in age and baseline creatinine between the two clusters were relatively modest.
Overall, the two-cluster model captures a broad division between a renal-predominant, lower-activity phenotype and a multi-system, high-activity phenotype.

The estimated cluster probability matrix $\hat{\pi}$ for the $2 \times 2$ model is given by
\[
\hat{\pi} =
\begin{bmatrix}
   0.235 & 0.099 \\
   0.456 & 0.210
\end{bmatrix},
\]
which reveals a dominant baseline cluster and a dominant longitudinal trajectory cluster. Approximately \(67\%\) of patients are assigned to the second baseline cluster, while nearly \(70\%\) belong to the first longitudinal cluster, which is characterized by relatively stable and lower creatinine values over time. The largest entry in $\hat{\pi}$, \(\hat{\pi}_{2,1} = 0.456\), corresponds to patients simultaneously belonging to the multi-view combination of $S_{\mathrm{im }} \times L_{\mathrm{s}}$, representing nearly half of the entire cohort. This indicates a strong but not exclusive alignment between the dominant baseline profile and the stable trajectory pattern.

\begin{table}[h]
\centering
\caption{Distribution of ESKD Outcomes and Berden Biopsy Classes in the 2×2 Cluster Configuration}
\begin{tabular}{lcccc}
\hline
\textbf{\textit{Cluster}} & \textbf{$L_{\mathrm{v}}\times S_{\mathrm{po}}$} & \textbf{$L_{\mathrm{v}}\times S_{\mathrm{im}}$} & \textbf{$L_{\mathrm{s}}\times S_{\mathrm{po}}$} & \textbf{$L_{\mathrm{s}}\times S_{\mathrm{im}}$} \\
\hline
Total (N)              & 62  & 24  & 125 & 59 \\
ESKD (n)               & 6   & 4   & 16  & 4  \\
ESKD (\%)              & 9.7 & 16.7 & 12.8 & 6.8 \\
Crescentic (n)         & 8   & 5   & 19  & 12 \\
Crescentic (\%)        & 12.9 & 20.8 & 15.2 & 20.3 \\
Focal (n)              & 12  & 8   & 32  & 16 \\
Focal (\%)             & 19.4 & 33.3 & 25.6 & 27.1 \\
Mixed (n)              & 14  & 2   & 20  & 7  \\
Mixed (\%)             & 22.6 & 8.3 & 16.0 & 11.9 \\
Sclerotic (n)          & 5   & 2   & 9   & 3  \\
Sclerotic (\%)         & 8.1 & 8.3 & 7.2 & 5.1 \\
Unknown Berden (n)   & 23 & 7 & 45 & 21 \\
Unknown Berden (\%)  & 37.1 & 29.2 & 16.8 & 35.6 \\
\hline
\end{tabular}
\end{table}

We further examine the relationship between the identified multi-view clusters, end-stage kidney disease (ESKD), and Berden renal biopsy classifications (focal, crescentic, mixed, and sclerotic) \citep{berden2010histopathologic}, as shown in Table 8. We then assessed whether the distribution of ESKD outcomes and Berden biopsy classes differed across the $2\times2$ cluster configuration. For ESKD, Fisher’s exact test was applied due to small cell counts, and no significant differences were observed across clusters ($p=0.501$). Similarly, the distribution of Berden biopsy classes does not differ significantly between clusters ($\chi^2 = 6.99$, $p=0.86$). These results indicate that, within the $2\times2$ latent configuration, histopathological severity and renal outcomes were not strongly associated with cluster membership.

\section{Discussion}
\label{sec_discussion}

The proposed multi-view mixture modeling framework offers a flexible, principled approach for jointly analyzing baseline clinical covariates and longitudinal biomarker trajectories. It can be applied to any setting involving joint clustering of fixed-dimensional features and longitudinal trajectories.

A first possible extension of this work is to move from a single longitudinal feature setting to a multivariate longitudinal setting by treating each biomarker as a separate longitudinal view within the same mixture framework. In the current framework, a single latent Neural ODE captures the temporal evolution of the longitudinal trajectory, which is jointly modeled with the static view through a two-dimensional cluster membership structure. This formulation can be generalized to multiple biomarkers by introducing multiple latent trajectories, one for each longitudinal biomarker, while coupling them through a shared latent mixture structure that links the static view and all longitudinal views. Under this construction, each biomarker is allowed to exhibit its own temporal patterns, while cluster membership remains informed by the joint evolution of all longitudinal trajectories and static features. 

A second potential extension concerns the observation model linking latent trajectories to observed measurements. In the current formulation, a Gaussian likelihood is adopted for simplicity and computational convenience. However, biomedical measurements often deviate from normality, exhibiting skewness, heavy-tailed behaviour, or heteroscedastic measurement error. The Gaussian assumption can be relaxed by substituting alternative likelihood specifications that more accurately reflect the underlying data-generating process.

\newpage
\appendix

\section{List of baseline variables for PARADISE project}
\renewcommand{\thetable}{A\arabic{table}}
\setcounter{table}{0}

\begin{table}[h]
    \caption{Description of Baseline Variables for the Irish ANCA cohort}
    \centering
    \begin{tabular}{|l|l|}
        \hline
        Variable Name & Description \\ 
        \hline
        Age   & The individuals age in years at the time of diagnosis   \\ 
        Creatinine   & The individuals serum creatinine at the time of diagnosis in $\mu$mol/L  \\ 
        CRP   & The individuals CRP at the time of diagnosis in mg/L   \\ 
        gender   & Patient Gender as Male = 1, Female = 2    \\ 
        Constitutional   & The presence of constitutional symptoms: No = 1, Yes = 2   \\ 
        Musculoskeletal   & The presence of musculoskeletal symptoms: No = 1, Yes = 2 \\
        Cutaneous   & The presence of cutaneous symptoms: No = 1, Yes = 2  \\ 
        Eye   & The presence of eye symptoms: No = 1, Yes = 2  \\ 
        Mucosa   & The presence of mucosa symptoms: No = 1, Yes = 2  \\ 
        ENT   & The presence of ear-nose-throat symptoms: No = 1, Yes = 2  \\ 
        Chest   & The presence of lung symptoms: No = 1, Yes = 2  \\ 
        Cardio   & The presence of cardiovascular symptoms: No = 1, Yes = 2  \\ 
        Abdominal   & The presence of abdominal symptoms: No = 1, Yes = 2  \\ 
        Kidney   & The presence of kidney symptoms: No = 1, Yes = 2  \\ 
        CNS   & The presence of central nervous system symptoms: No = 1, Yes = 2  \\ 
        PNS   & The presence of peripheral nervous system symptoms: No = 1, Yes = 2  \\ 
        ANCA   & The ANCA type: ANCA negative = 1, MPO-/p-positive = 2, PR3/c–positive = 3  \\ 
        \hline
    \end{tabular}
    \label{tab:basic_table}
\end{table}
\label{app:appendix_label_1}

\newpage
\section{Baseline Phenotypic Characteristics}
\renewcommand{\thetable}{B\arabic{table}}
\setcounter{table}{0}
\begin{table}[h]
\caption{Distribution of Baseline Static Features Across Patients in PARADISE Cohort }
\centering
\begin{tabular}{l l}
\hline
\textbf{Characteristic} & \textbf{Overall (N = 282)} \\
\hline
\textbf{Demographics} & \\
Age, years (mean $\pm$ SD) & 61.3 $\pm$ 13.5 \\
Creatinine, $\mu$mol/L (mean $\pm$ SD) & 186.6 $\pm$ 139.9 \\
CRP, mg/L (mean $\pm$ SD) & 57.4 $\pm$ 70.2 \\
Male, n (\%) & 161 (57.1\%) \\
Female, n (\%) & 121 (42.9\%) \\
\hline
\textbf{Clinical Manifestations}, n (\%) & \\
Constitutional symptoms & 83 (29.4\%) \\
Musculoskeletal & 103 (36.5\%) \\
Cutaneous & 40 (14.2\%) \\
Eye involvement & 20 (7.1\%) \\
Mucosal involvement & 12 (4.3\%) \\
ENT involvement & 82 (29.1\%) \\
Chest involvement & 101 (35.8\%) \\
Cardiovascular involvement & 5 (1.8\%) \\
Abdominal involvement & 9 (3.2\%) \\
Kidney involvement & 177 (62.8\%) \\
CNS involvement & 4 (1.4\%) \\
Peripheral nervous system & 18 (6.4\%) \\
\hline
\textbf{ANCA subtype}, n (\%) & \\
PR3--ANCA & 138 (48.9\%) \\
MPO--ANCA & 136 (48.2\%) \\
ELISA negative/other & 8 (2.8\%) \\
\hline
\end{tabular}
\end{table}
\label{app:appendix_label_2}

\begin{table}[h]
\caption{Cluster-level summary statistics for the 2\(\times\)2 fixed-dimensional configuration (N = 282). Organ system involvement is shown as the proportion of patients in each cluster. Continuous variables are summarized using median, mean, standard deviation (SD), and range.}
\centering
\small
\begin{tabular}{lcc}
\hline
\textbf{Variable} & \textbf{Cluster 1} & \textbf{Cluster 2} \\
\hline
N & 196 & 86 \\
Constitutional (\%) & 11.73 & 69.77 \\
Musculoskeletal (\%) & 18.88 & 76.74 \\
Cutaneous (\%) & 7.14 & 30.23 \\
Eye (\%) & 3.57 & 15.12 \\
Mucosa (\%) & 2.04 & 9.30 \\
ENT (\%) & 23.47 & 41.86 \\
Chest (\%) & 20.92 & 69.77 \\
Cardio (\%) & 1.53 & 2.33 \\
Abdominal (\%) & 0.51 & 9.30 \\
Kidney (\%) & 47.96 & 96.51 \\
CNS (\%) & 2.04 & 0.00 \\
PNS (\%) & 7.14 & 4.65 \\
Female (\%) & 45.92 & 36.05 \\
MPO-ANCA (\%) & 51.02 & 41.86 \\
PR3-ANCA (\%) & 44.90 & 58.14 \\
Age, Median (years) & 63.5 & 59.0 \\
Age, Mean (years) & 62.77 & 58.01 \\
Age, SD & 13.13 & 13.93 \\
Age, Min--Max & 21--93 & 24--84 \\
Creatinine, Median (\(\mu\)mol/L) & 152.0 & 140.0 \\
Creatinine, Mean (\(\mu\)mol/L) & 191.40 & 175.59 \\
Creatinine, SD & 148.68 & 117.38 \\
Creatinine, Min--Max & 35--992 & 44--791 \\
CRP, Median (mg/L) & 16.2 & 53.5 \\
CRP, Mean (mg/L) & 46.09 & 83.28 \\
CRP, SD & 61.85 & 80.90 \\
CRP, Min--Max & 0--309 & 1--290 \\
\hline
\end{tabular}
\label{tab:cluster_summary_2x2}
\end{table}




\end{document}